\documentclass{mn2e}
\usepackage{times,graphicx,amssymb,color}

\title[Inner edge of the Oort cloud]{Reassessing the formation of the Inner Oort cloud in an embedded star cluster II: Probing the
inner edge}
\author[R. Brasser and M. E. Schwamb]{R. Brasser$^{1,2}$ and M. E. Schwamb$^1$\\
$^1$Institute for Astronomy and Astrophysics, Academia Sinica; 11F AS/NTU, National Taiwan University, 1 Roosevelt Rd., Sec. 4,
Taipei 10617, Taiwan\\
$^2$ Earth-Life Science Institute, Tokyo Institute of Technology, Meguro, Tokyo 152-8551, Japan}

\begin{document}
\maketitle
\begin{abstract}
The detached object Sedna is likely at the inner edge of the Oort cloud, more precisely the inner Oort cloud (IOC). Until recently it
was the sole member of this population. The recent discovery of the detached object 2012 VP113 has confirmed that there should be more
objects in this region. Three additional IOC candidates with orbits much closer to Neptune have been proposed in the past decade since
Sedna's discovery: 2000 CR105, 2004 VN112 and 2010 GB174. Sedna and 2012 VP113 have perhelia near 80~AU and semi-major axes over
250~AU. The latter three have perihelia between 44~AU and 50~AU and semi-major axes between 200~AU and 400~AU. Here we determine
whether the latter three objects belong to the IOC or are from the Kuiper Belt's Extended Scattered Disc (ESD) using numerical
simulations. We assume that the IOC was formed when the Sun was in its birth cluster. We analyse the evolution of the IOC and the
Scattered Disc (SD) during an episode of late giant planet migration. We examine the impact of giant planet migration in the context of four
and five planets. We report that the detached objects 2004 VN112 and 2010 GB174 are likely members of the IOC that were placed there
while the Sun was in its birth cluster or during an episode of Solar migration in the Galaxy. The origin of 2000 CR105 is ambiguous 
but it is likely it belongs to the ESD. Based on our simulations we find that the maximum perihelion distance of SD objects is 41~AU 
when the semi-major axis is higher than 250~AU. Objects closer in are subject to mean motion resonances with Neptune that may 
raise their perihelia. The five planet model yields the same outcome. We impose a conservative limit and state that all objects with 
perihelion distance $q>45$~AU and semi-major axis $a>250$~AU belong to the inner Oort cloud.
\end{abstract}

\section{Introduction}
In 2003 Brown et al. (2004) discovered distant Solar System object Sedna whose orbit defied all dynamical models that existed at that
time. Sedna resides in an area of phase space that was previously thought to have been unoccupied. Effectively its orbit is dynamically
isolated from the rest of the Solar System. With a semi-major axis of only 533~AU, its orbit is nearly two orders of magnitude smaller
than that of a typical Oort cloud object, but at the same time is also much larger than that of most Kuiper belt objects (KBOs). With a
perihelion distance of 76~AU well beyond the orbit of Neptune, dynamical interaction between Neptune and Sedna is limited to precession.
With an aphelion of $\sim$1000 AU, Sedna is too close to the Sun to feel the effects of passing stars and galactic tides in the 
present-day solar neighbourhood that act on classical Oort cloud objects. The only viable mechanism to place Sedna on its current 
orbit is through a gravitational force that was active at some time earlier in the history of the Solar System. 

Several formation scenarios have been put forth to explain Sedna's extreme orbit and the population of objects on similar orbits
predicted to exist as a result (for the purposes of this paper we refer to this population as the Inner Oort cloud; IOC): an encounter
of a solar mass star passing at about 500-1000~AU from the Sun at low velocity (Morbidelli \& Levison 2004, Keyon \& Bromley 2004),
gravitational scattering by a distant planetary-sized solar companion (Gomes et al. 2006, Gladman \& Chan, 2006), solar migration in
the Galaxy during an episode of intense planetesimal scattering (Kaib et al. 2011), gravitational scattering produced by multiple
stellar encounters while the Sun was residing in an embedded cluster (Brasser et al. 2006, Brasser et al. 2012) or an open cluster
(Kaib \& Quinn, 2008). With Sedna alone it is impossible to distinguish between the formation mechanisms. In the past decade, it has
been difficult to further refine the Sedna formation models and the dynamical history of the outer Solar System because no other
objects in the IOC region with perihelia greater than 70 AU, had been found. The situation has recently changed with the discovery of
2012 VP113, a distant object with perihelion distance $q=$80~AU and semi-major axis of only $a=$270~AU (Trujillo \& Sheppard, 2014).
Like Sedna, 2012 VP113 does not dynamically interact with Neptune and the same arguments pertain to its origin as those of Sedna.
Though the discovery of 2012 VP113 confirms the existence of the IOC, with only two confirmed members it is still not possible to 
uniquely identify the IOC formation mechanism.

Dynamically the orbits of 2012 VP113 and Sedna have a clear origin distinct from the Kuiper belt and Scattered Disc (SD; Duncan \&
Levison, 1997) as part of the IOC, but what about the formation of other distant bodies closer to Sun than Sedna and 2012 VP113 with
perihelia detached from Neptune often referred to as `detached' KBOs (Gladman et al 2008)? These objects reside on orbits with
typically much lower semi-major axes ($\sim$225-400~AU) and perihelia (between 44~AU and 49~AU) than Sedna and 2012 VP113 (Gladman et
al. 2002). For this paper we define this detached population as orbits with semi-major axis $a>50$~AU and $q>40$~AU. The orbital
elements of all known detached KBOs to date are listed in Table~\ref{iocesd}. Do these detached KBOs also belong to the IOC or are 
they instead emplaced by Neptune as part of the Kuiper belt's Extended Scattered Disc (ESD; Gladman et al. 2002)? We define an ESD 
object as a detached KBO not in resonance with Neptune. At present it is unclear where the boundary lies between the ESD and the IOC. 

Dynamical simulations by Gomes et al. (2005) and Lykawka \& Mukai (2007) show that high-order mean motion resonances with Neptune, 
coupled with the Kozai mechanism (Kozai, 1962) in these resonances, can produce an ESD. Objects can become permanently trapped in the
ESD during Neptune's outward migration when they reach low eccentricity because they pop out of the resonance, thereafter no longer
interacting with Neptune (Gomes, 2011). Gomes et al. (2005) and Lykawka \& Mukai (2007) show that the combined resonance-Kozai 
mechanism operates for objects with semi-major axis $a\lesssim 200-250$~AU, and the maximum perihelion distance is typically 60~AU. 
Therefore it is most likely that any object with $50<a\lesssim200$~AU and $40<q<60$~AU belongs to the ESD, but objects with semi-major 
axis $a \gtrsim 200$~AU and perihelion distance $q\gtrsim 60$~AU could have a different dynamical origin and are most likely part of 
the IOC in which both Sedna and 2012 VP113 belong. Objects with $a\lesssim200$~AU and $q>60$~AU would also be classified as belonging 
to the IOC. From these criteria the origin of  2000 CR105 (Gladman et al., 2002), 2004 VN112 (Becker et al., 2008), and 2010 GB174 
(Chen et al., 2013)  is less clear: they could be part of the IOC i.e. they were placed on their current orbits during the early stages 
of the Solar System and evolved very little since, or they could be part of the ESD and were placed there during giant planet 
migration when they interacted with Neptune.

The current studies presented above are based on the assumption there were only four giant planets (two gas giants and two ice
giants) present in the outer Solar System during planetesimal driven migration. Recent modelling (Batygin et al. 2010, Nesvorn\'{y} 2011,
Nesvorn\'{y} \& Morbidelli 2012) find that a five-planet scenario (two gas giant planets and three ice giants), where one of the ice
giants is subsequently ejected by Jupiter, can also reproduce the present day orbits of the giant planets in the Solar System. In the
past several years, there has been growing evidence that the five-planet scenario may provide a better match to the architecture of the
outer Solar System (Batygin et al. 2012, Nesvorn\'{y} \& Morbidelli 2012, Brasser \& Lee 2014). During the ejection process, the third
ice giant may have spent on the order of a few tens of thousands of years on an eccentric orbit with a semi-major axis greater than
50~AU before finally exiting the Solar System. Although this time is relatively short compared to the age of the Solar System, it may
still be sufficient for secular resonances to act on the orbits of SD objects and IOC objects. To date no study has been performed
that determined the influence of the fifth planet on the SD and IOC.

The discovery of 2012 VP113  has fostered a need to re-examine the distinction between the IOC and the ESD and the boundary where these
two populations are distinguishable in terms of the cluster birth scenario. This re-examination should occur in the framework of
both four and five planet models. Though the embedded cluster IOC model does not present a unique solution to Sedna's origin, it does
predict a unique series of orbital distributions that can be tested and constrained by observations. The majority of the other proposed
scenarios have a rather low probability of occurrence, while most stars form in embedded star clusters (Lada \& Lada, 2003).
Additionally, with the recent discovery of a possible solar sibling (Ram\'{i}rez et al., 2014), the star cluster birth becomes more
viable. In this paper we aim to quantify which objects belong to the IOC and which ones are part of the ESD, issue a few criteria
for observers to immediately classify an object as belonging to the IOC, identify which objects serve as a crucial test for the 
embedded cluster model of Sedna's formation, and simultaneously focus recovery efforts on these objects. We shall focus mostly on 
using the established four planet models of Levison et al. (2008) and Brasser \& Morbidelli (2013) and then examine the influence of 
the fifth planet.

This paper is organised as follows. In the Section 2 we summarize the embedded cluster birth formation model for the IOC. Section
3 we present the methodology of our numerical simulations for the evolution of the SD and the IOC as it formed during the stellar
cluster era. This is followed by the results of these simulations described in Section 4. In Section 5 we explore how a fifth giant
planet could have sculpted the outer Solar System. We present our conclusions in the last section.
\begin{table}
\begin{tabular}{c|cccc}
  Object & Semi-major axis & Perihelion distance & Inclination\\ \hline \hline
 \multicolumn{4}{c}{Confirmed IOC Members } \\
  2012 VP113 & 268 & 80.3 & 24$^\circ$ \\
  (90377) Sedna & 533 & 76.2 & 12$^\circ$ \\ \hline
   \multicolumn{4}{c}{IOC candidates } \\
  2010 GB174 & 368 & 48.7 & 22$^\circ$ \\
  2004 VN112 & 334 & 47.3 & 27$^\circ$\\
  2000 CR105 & 230 & 44.2 & 23$^\circ$ \\ \hline
    \multicolumn{4}{c}{Detached KBOs } \\
  2004 XR190 & 57.8 & 51.4 & 47$^\circ$ \\
  2005 TB190 & 75.4 & 46.2 & 27$^\circ$\\
  2009 KX36 & 67.8 & 44.7 & 23$^\circ$ \\
  2004 PD112 & 64.3 & 43.6 & 7$^\circ$\\
  2008 ST291 & 99.1 & 42.5 & 21$^\circ$\\
  2002 CP154 & 52.6 & 42.0 & 2$^\circ$\\
  2006 AO101 & 52.9 & 41.9 & 1$^\circ$\\
  2000 YW134 & 58.3 & 41.2 & 20$^\circ$\\
  2005 EO197 & 63.3 & 41.1 & 26$^\circ$\\
  2005 CG81 & 54.2 & 41.1 & 26$^\circ$\\
  1995 TL8 & 52.6 & 40.2 & 0$^\circ$\\
  2010 ER65 & 99.2 & 40.0 & 21$^\circ$
\end{tabular}
\caption{Semi-major axis, perihelion distance and inclination of the two IOC objects, IOC candidates and the known detached KBOs 
(semi-major axis $a>50$~AU and $q>40$~AU) per July 2014. Elements taken from the Minor Planet Centre
(http://minorplanetcenter.net/iau/Ephemerides/Distant/index.html) on July 1 2014.}. 
\label{iocesd}
\end{table}

\section{Embedded Cluster Birth}
Here we provide a brief overview and history of the embedded cluster birth model. Sedna's proximity to the Sun and detached orbit led
to the suggestions that it could be part of the Inner Oort cloud (IOC) (Hills, 1981). This suggestion led Brasser et al. (2006) to
study the origin of Sedna while the Sun was still in its embedded birth cluster. Their motivation rested on the suggestion that most
stars form in embedded star clusters (Lada \& Lada, 2003). More recently Ram\'{i}rez et al.(2014) have spectroscopically identified the
first candidate solar sibling strengthening the argument that the Sun was likely born in a stellar cluster. Brasser et al. (2006)
performed numerical simulations of Jupiter and Saturn scattering planetesimals in their vicinity, which at large distances were then
subjected to stellar encounters and tidal forces from the cluster. They concluded that Sedna would be at the inner edge of the Oort
cloud if the central density of the cluster peaked at 10\,000 solar masses per cubic parsec ($M_\odot$~pc$^{-3}$). Schwamb et al.
(2010) compared their Palomar observations of distant Solar System objects with perihelia $q>50$~AU, and compared them to the models of
Brasser et al. (2006).  They only considered objects with $q>50$~AU to minimise the probability of dynamical interaction with
Neptune. They concluded that the cluster with central density 10\,000 $M_\odot$~pc$^{-3}$ was the most consistent with
their data, where Sedna was situated close to the inner edge of the IOC, ruling out with high confidence a model with central density
10$^5$~$M_\odot$~pc$^{-3}$ (where Sedna was closer to the central part of the IOC).

Even though the Brasser et al. (2006) models demonstrated the feasibility of producing Sedna and the IOC while the Sun was in its birth
cluster, it contained several approximations and dynamical shortcomings. Employing a much more sophisticated computer code and a more
realistic model and initial conditions for the star cluster, Brasser et al. (2012) reassessed the formation of the Inner Oort cloud in
an embedded star cluster. They used two types of central potential for the cluster: the Hernquist potential (Hernquist, 1990) and the
Plummer potential (Plummer, 1911). The former is more centrally condensed than the other. They also varied the number of stars in each
cluster, ranging from 50 to 1000, the central concentration, and embedded gas lifetime. Table~\ref{cluster} presents a summary of the
various parameters and their values. Brasser et al. (2012) concluded that the production of Sedna was a generic outcome and that all
cluster models that were tested in that work fit the observations of Schwamb et al. (2010). Trujillo \& Sheppard (2014) combine their
survey results with that of Schwamb et al (2010), testing an orbital distribution similar to that produced by the  Brasser et al.
(2006) and Brasser et al. (2012) cluster birth scenario. They find that the discovery of Sedna and 2012 VP113 is consistent with
stellar cluster produced population if Sedna and 2012 VP113 are at the starting edge of the orbital distribution. However, the
simulations of Brasser et al. (2006) and Brasser et al. (2012) stopped when the Sun left the cluster. The subsequent evolution was not
taken into account and is thus done here.

\begin{table}
\begin{tabular}{c|c}
  Parameter &  Value\\ \hline \hline
Potential & Hernquist or Plummer \\
Concentration & 3 or 6 \\
Gas lifetime & 2~Myr or 4~Myr \\
Number of stars & 50, 100, 250, 350, 500, 750, 1000\\
\end{tabular}
\caption{Cluster parameters and their values used in the IOC formation simulations of Brasser et al. (2012).}
\label{cluster}
\end{table}

\section{Methods: Numerical simulations using the four planet case}
\label{sec:4planet}
At present it is unclear what is the extent of the ESD, and, assuming a cluster birth scenario, where the boundary lies between the ESD
and the IOC. To answer this we need to know how the dynamical evolution of the IOC and the SD shaped their current populations. We
use numerical simulations subjecting the IOC and SD to giant planet migration and evolve the populations over the remaining age of the
Solar System including the effects of passing stars and galactic tides in the current solar environment. 

\subsection{Giant Planet Migration}
\label{sec:migration} 
\begin{figure}
\resizebox{\hsize}{!}{\includegraphics[angle=-90]{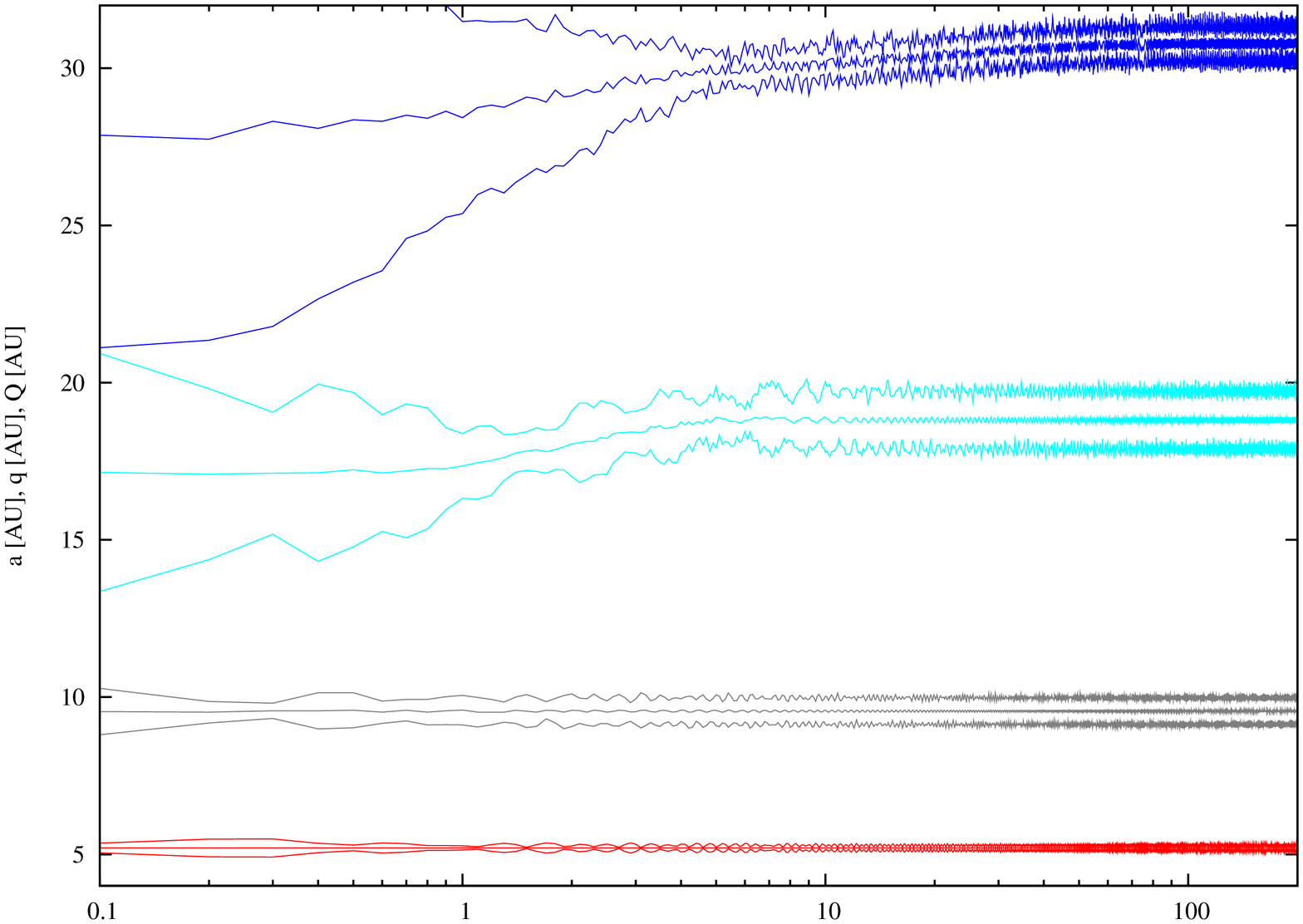}}
\caption{The semi-major axes, perihelia and aphelia of the giant planets for the simulation with four planets. This is identical to 
Run A of Levison et al. (2008). The red lines correspond to Jupiter, the purple ones to Saturn, cyan for Uranus and deep blue for 
Neptune.}
\label{evo5}
\end{figure}

For direct comparison, we make use of the same four-planet migration history for both the IOC and SD simulations. We emphasise that we
do not distinguish between early or late giant planet migration, but only use the fact that the giant planets did migrate. We have used
the recipe for giant planet evolution described in  Levison et al. (2008). More precisely, we re-enact the evolution shown in their Run
A in which, at the end of the phase of mutual close encounters among the planets, Neptune's semi-major axis is $a_N = 27.5$~AU and its
eccentricity is $e_N = 0.3$. Uranus's semi-major axis and eccentricity are $a_U = 17.5$~AU and $e_U = 0.2$. The mutual inclination of
both planets is approximately 1$^\circ$. Jupiter remains at 5~AU, Saturn stays at 9.5~AU. Uranus migrates from 17~AU to nearly 19~AU
while Neptune migrates outwards to settle close to 31~AU. The evolution of the planets is shown in Fig.~\ref{evo5}. We want to 
emphasize that the real evolution of the planets cannot be traced, so that we do not expect that the evolution we consider is exactly 
right. However, the evolution of Run A leads to final planetary orbits very similar to the current ones and shows a high compatibility 
with the currently known orbital structure of the Kuiper Belt (Levison et al., 2008). Hence we argue the evolution above is 
representative of what could have happened in reality, though with the caveat that this case was most likely a maximum planetary 
eccentricity case and therefore more likely to emplace detached objects further out from Neptune’s current orbit.

\subsection{Inner Oort cloud}
\label{sec:IOC}
For the IOC, we took as initial conditions the orbits of IOC objects from Brasser et al. (2012). We combined the end results for all
number of stars but we kept the other parameters fixed. These orbits of the IOC objects are taken at the epoch the Sun left the cluster
and thus they underwent no subsequent evolution. We performed sixteen simulations, eight for IOCs formed during star clusters modelled
using the more centrally concentrated Hernquist potential (Hernquist, 1990), and another eight using the less concentrated Plummer
models (Plummer, 1911). The different runs for each potential are for various values of the central concentration and gas removal time.
We lumped together the IOCs for different number of stars in the cluster since the structure of the IOC appeared insensitive to the
cluster size (Brasser et al., 2012). This resulted in sixteen different simulations (2 potentials, 2 concentrations and 2 gas
lifetimes). 

To determine where the inner edge of the IOC resides we need to subject the IOC to giant planet migration. The outward migration of
Neptune, together with a possible high-eccentricity phase, could eliminate any IOC objects with perihelia up to about 40~AU, and
possibly higher. These simulations are intended to determine the minimum semi-major axis and perihelion distance where we can safely
argue the object is part of the IOC rather than the ESD.

The IOC was composed of massless test particles whose orbital elements were used directly after the Sun escaped the star cluster. The
system of planets and test particles were numerically integrated for 200~Myr with SWIFT RMVS4 (Levison \& Duncan, 1994).
We included the influence of the Galactic tide and passing stars. The tides were implemented using the method of Levison et al. (2001)
with a Galactic density of 0.1~$M_{\odot}$~pc$^{-3}$ (Holmberg \& Flynn, 2000) and Galactic rotational velocity 30.5
km~s$^{-1}$~kpc$^{-1}$ (McMillan \& Binney, 2010). The perturbations from passing stars were included as described in Heisler et al.
(1987) with the stellar spectral data and velocity of Garcia-Sanchez et al. (2001). The time step was 0.4~yr. Objects were removed
from the simulation if they were farther than 2\,000~AU from the Sun, were unbound, hit a planet or came closer than 0.5~AU from the
Sun. We deliberately removed any test particles farther than 2\,000~AU from the Sun because we are interested in the evolution of the
innermost part of the IOC, where 2012 VP113 and Sedna reside, and where the Galactic tides are almost inactive. We found that
particles farther than 2\,000~AU underwent substantial evolution in their perihelion distance because of the Galactic tide and they
could contaminate the inner core of the IOC. This cutoff is somewhat closer than the 3\,000~AU used by Brasser \& Morbidelli (2013) 
but the results are virtually independent of it.

After the migration of the planets, as described in Section \ref{sec:migration}, we continued our simulations for another 4~Gyr. We
kept the planets on their final orbits just after migration. This second set of simulations were performed with SCATR (Kaib et al.,
2011), which is a Symplectically-CorrectedAdaptive Timestepping Routine. It is based on SWIFT's RMVS3 (Levison \& Duncan, 1994). It has
a speed advantage over SWIFT's RMVS3 or MERCURY (Chambers, 1999) for objects far away from both the Sun and the planets where the time
step is increased. We set the boundary between the regions with short and long time step at 300~AU from the Sun (Kaib et al. 2011).
Closer than this distance the computations are performed in the heliocentric frame, like SWIFT's RMVS3, with a time step of 0.4~yr.
Farther than 300~AU, the calculations are performed in the barycentric frame and we increased the time step to 50~yr. The error in the
energy and angular momentum that is incurred every time an object crosses the boundary at 300~AU is significantly reduced through the
use of symplectic correctors (Wisdom et al., 1996). For the parameters we consider, the cumulative error in energy and angular momentum
incurred over the age of the solar system is of the same order or smaller than that of SWIFT's RMVS3. The same Galactic and stellar
parameters as in the first set of simulations were used, and the same removal conditions.

\subsection{Extended Scattered Disc}
\label{sec:4planetESD}
For the creation, and subsequent evolution, of the ESD we did not perform any new Scattered Disc simulations. We use the same 
definition of an SD object as Brasser \& Morbidelli (2013): an SD object is defined to have $a < 1\,000$ ~AU and $q >$ 30~A. We 
examined the resulting orbital distributions produced in the SD simulations performed by Brasser \& Morbidelli (2013). The Brasser \& 
Morbidelli (2013) simulations surpass those of Gomes et al. (2005) and Lykawka \& Mukai (2007) in that they use a controlled migration 
evolution of the giant planets and contain many more test particles. We briefly summarise below the process Brasser \& Morbidelli 
(2013) used to generate their SD and ESD populations.

Brasser \& Morbidelli (2013) use the same initial conditions and giant planet migration as described in Section \ref{sec:migration}.
After the planets had stopped migrating the remaining test particles were integrated with SWIFT RMVS3 for another 3.8~Gyr. They took
the planets and comets at the end of the planetary migration phase and removed all comets that were further than 3\,000~AU from the
Sun. To correctly simulate the evolution of the SD, and that of the Centaurs and Jupiter-family comets, the planets needed to be on
their current orbits, or match these as closely as possible. At the end of their migration simulations Uranus was too close to the Sun.
Therefore after the migration was completed, Brasser \& Morbidelli (2013) ran a second migration stage where they artificially migrated
Uranus outwards by 0.25~AU to its current orbit over a time scale of 5~Myr while Neptune remained in place. The final configuration
matched the current positions and secular properties of the giant planets. For the IOC simulations described in Section
\label{sec:IOC}, migrating Uranus to its current position was not needed since Uranus' influence is negligible on the orbits of even
the closest IOC orbits. Thus this extra piece of migration was skipped.

After this second stage of migration the planets and SD objects were integrated for 4~Gyr. Brasser \& Morbidelli (2013) cloned
remaining objects immediately after this second stage of giant planet migration, and again after 1~Gyr and after 3.5~Gyr, to keep
enough objects alive for a variety of statistical purposes. Cloning was achieved by adding a random deviation of $10^{-6}$ radians to
the comets' mean anomaly, keeping all the other elements fixed. The time step was 0.4~yr. Objects were removed from the simulation if
they strayed further than 3\,000~AU from the Sun, or hit the Sun or a planet.

After 4~Gyr the SD in the simulations of Brasser \& Moribdelli (2013) typically contained 2\,000 comets. Their typical efficiency of
implanting material in the SD after 4~Gyr was 0.6\%, implying that they effectively integrated 280\,000 test particles per simulation,
or about 1 million test particles in total. We shall make use of this in the next section.

\section{Results for the four planet case}
In this section the results from our numerical simulations are presented.

\subsection{Extended Scattered Disc}
\begin{figure}
\resizebox{\hsize}{!}{\includegraphics[angle=-90]{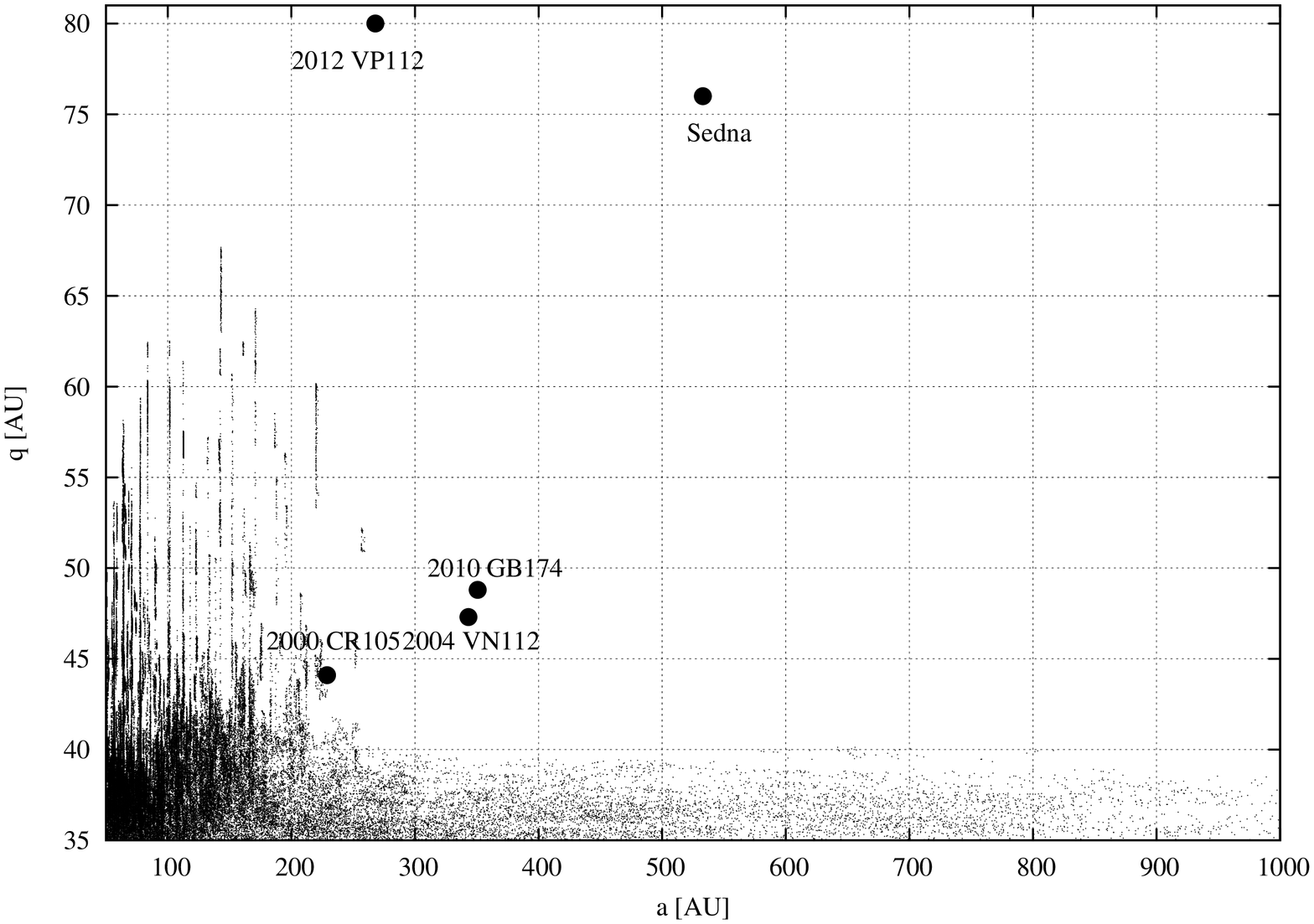}}
\caption{Orbits of synthetic SD and ESD particles  emplaced after 4-giant planet migration from the models of Brasser \& Moribdelli 
(2013) integrated over the age of the Solar System.  Larger filled circles plot the 5 know detached KBOs with $a \gtrsim 200$~AU and
perihelion distance $q\gtrsim 40$~AU }
\label{sd}
\end{figure}

We determined the extent and orbital structure of the  SD and ESD with the help of Fig.~\ref{sd}. The dots are the
time-evolution in semi-major axis-perihelion space of all objects in the SD with $q>35$~AU and $a>50$~AU over the last 400~Myr of the
simulation. The bullets depict the location of 2000 CR105, 2004 VN112, 2010 GB174, Sedna and 2012 VP113. 
It is clear that there are three distinct regions determined by the presence or absence of objects. These regions are:
\begin{itemize}
 \item $a<250$~AU and $q<60$~AU
 \item $a>250$~AU and $q<40$~AU
 \item $a>250$~AU and $q>40$~AU
\end{itemize}
We shall discuss each of these below.

In the first region there are a multitude of distinct narrow regions where objects can reach high perihelion at constant semi-major
axis. These peaks correspond to mean-motion resonances with Neptune coupled with the Kozai effect (Gomes et al., 2005; Lykawka \&
Mukai, 2007). Objects get trapped in resonance with Neptune as the latter migrates outwards, and once the particles' eccentricity
decreases significantly they are released from the resonance and get stranded at high perihelion. This mechanism only operates up until
$a \sim 250$~AU, roughly corresponding to the 1:24 resonance with Neptune, in agreement with Gomes et al., 2005 and Lykawka \& Mukai,
2007. At larger semi-major axis the resonance trapping mechanism is ineffective (as evident by the lack of peaks at higher semi-major
axis).

The second region is the SD. It appears planet migration and dynamical interaction beyond 250~AU does not produce objects
with $q>40$~AU. The sharp upper limit in perihelion distance of the SD at large semi-major axis not entirely expected. Earlier works
(Gomes, 2011; Gomes et al., 2005; Lykawka \& Mukai, 2007) often migrated Neptune to its current position on a nearly circular orbit,
without it going through a high-eccentricity phase. Our evolution of the giant planets has Neptune evolving through an eccentric phase
for several million years. An eccentric planet disturbs small bodies in its vicinity much more efficiently because of secular
eccentricity forcing. Unfortunately this same forcing damps the eccentricity of the planet on a short time scale. Since Neptune's
aphelion was temporarily beyond 40~AU one would expect a signature of this to be evident in the SD, but this does not appear to be so.
There are several reasons this could be the case. The first is that the high eccentricity phase of Neptune lasts too short to scatter
enough material out beyond 250~AU at high aphelion. Second, the eccentricity of Neptune was not high enough to detach objects beyond
40~AU. Thus the structure of the SD and the sharp upper perihelion limit of 40~AU appear to be a generic outcome after Neptune's
migration at low or moderate eccentricity. A very high eccentricity phase is ruled out by the current orbital structure of the Kuiper
belt (Batygin et al., 2011) and such high eccentricity evolution of Neptune is seldom seen in giant planet migration simulations
(Brasser \& Lee, 2014).

That brings us to the third region, $a>250$~AU and $q>40$~AU, which, in this plot, is as good as empty. Giant planet migration and 
interaction with Neptune does not fill this region as discussed above, so any objects that are found here must be part of the IOC. This
includes 2004 VN112 and 2010 GB174. But what about 2000 CR105? From the plot it is clear that 2000 CR105 is near a mean-motion
resonances with Neptune (the 1:21) and thus it is possible that this object got placed on its current orbit during the epoch of giant
planet migration, but it may also have been placed there early on by a close stellar encounter if the cluster model is correct. From
the simulations of Brasser \& Morbidelli (2013) its origin remains inconclusive: it could be an ESD object or IOC object.

\begin{figure}
\resizebox{\hsize}{!}{\includegraphics[angle=-90]{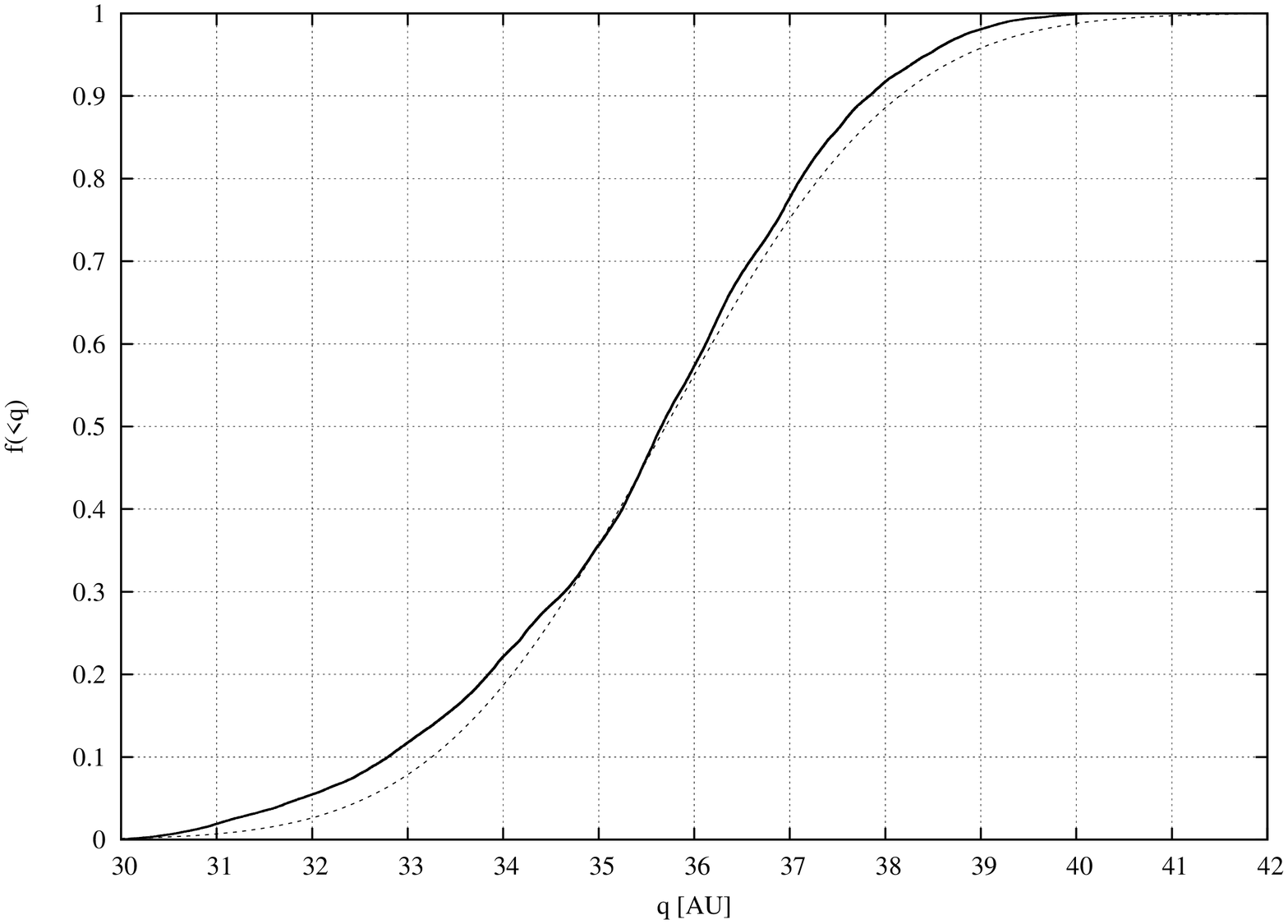}}
\caption{Cumulative distribution (solid line) of the perihelion distance of Scattered Disc objects with semi-major axis $a>250$~AU.
The dashed line is a best fit: a Gaussian with a mean $\mu_q = 35.7$~AU and standard deviation $\sigma_q = 1.9$~AU.}
\label{cumq}
\end{figure}

One may argue that the origin of 2004 VN112 and 2010 GB174 could be equally inconclusive, and that these objects could also have an ESD
origin, but that the simulations of Brasser \& Morbidelli (2013) did not have enough resolution to determine their origin. For
objects with $a>250$~AU the steady-state perihelion distribution for SD objects after 4~Gyr of evolution is approximately Gaussian with
mean $\mu_q = 35.7$~AU and standard deviation $\sigma_q = 1.9$~AU and thus it is predicted that only 0.04\% of objects have $q>42$~AU,
though we observed none. The cumulative distribution is displayed in Fig.~\ref{cumq}.
 We tried to determine whether 2004 VN112 and 2010 GB174 could have an ESD origin i.e. is it possible for SD objects to be pulled into
resonances with Neptune at $a \sim 350$~AU that raise their perihelion to 48~AU? Since the simulations of Brasser \& Morbidelli (2013)
did not have the resolution to determine whether the combined effect of resonance trapping and Kozai mechanism could operate at even
larger semi-major axes, we ran an extra set of simulations, in which we placed 36\,000 test particles on orbits with semi-major axis
uniformly between 250~AU and 500~AU and perihelion uniformly between 38~AU and 41~AU and inclination uniformly from 0 to 10$^\circ$.
With this setup we have an order of magnitude more objects with $a>250$~AU and $q>39$~AU than the simulations of Brasser \& Morbidelli
(2013). We ran these objects for 4~Gyr with SCATR with the giant planets on their current orbits, removing objects when they were
farther than 2\,000~AU from the Sun, hit a planet or came closer than 0.5~AU from the Sun. Ultimately we found no cases of objects
attaining orbits with $q>45$~AU and $a>250$~AU. We did, however, discover a different pathway to create ESD objects with $q<45$~AU that
we shall now discuss.
\begin{figure}
\resizebox{\hsize}{!}{\includegraphics[angle=-90]{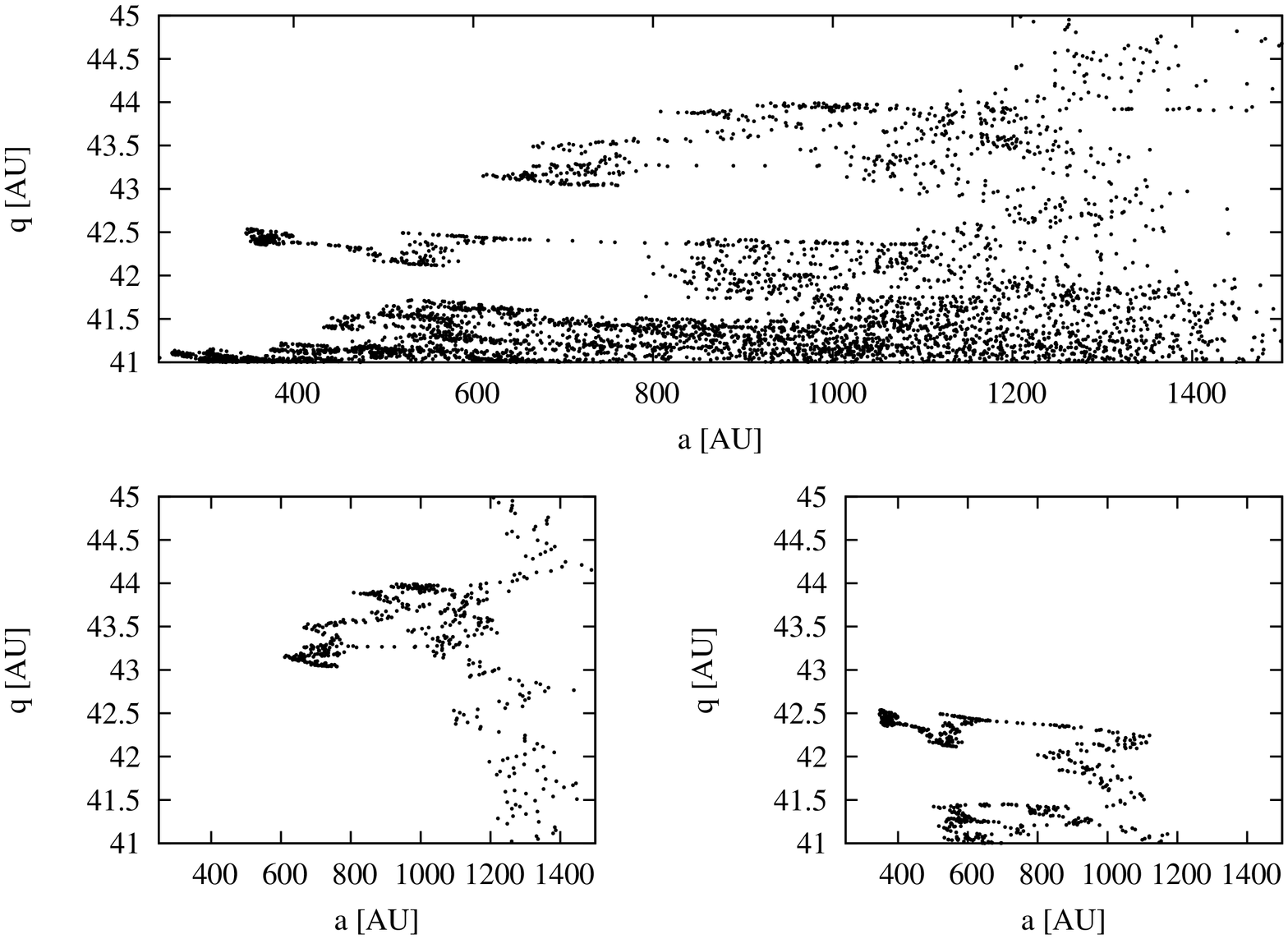}}
\caption{Top panel: Time evolution of the semi-major axis and perihelion of ESD objects with $q>41$~AU. The bottom two panels depict 
the individual evolution of two cases. The output interval is 1~Myr.}
\label{newpath}
\end{figure}
The top panel of Fig.~\ref{newpath} shows the time-evolution in semi-major axis-perihelion space of all objects in the ESD with
$q>41$~AU and $a>250$~AU over the entire 4~Gyr of one simulation. It is clear that some objects reach $q>42$~AU when $a\gtrsim 1000$~AU
but there are a few $a\lesssim 800$~AU. All of these objects have become detached from Neptune through the action of the Galactic tides
but some have diffused back to a lower semi-major axis. The evolution of two of these particles are shown on the bottom two panels. One
ends at $a \sim 600$~AU with $q \sim 43$~AU and another at $a \sim 400$~AU with $q \sim 42$~AU. However, we found no objects with
$q>45$~AU.

In summary, within this planetary migration framework our simulations and those of Gomes et al. (2005), Lykawka \& Mukai (2007) and
Brasser \& Morbidelli (2013) indicate that the upper limit of $q\sim40$~AU for SD objects with $q>250$~AU is real. Therefore, we
conclude that it is highly unlikely that 2004 VN112 and 2010 GB174 are ESD objects and that they have an IOC origin.

\subsection{Inner Oort cloud}
\begin{figure}
\resizebox{\hsize}{!}{\includegraphics[angle=-90]{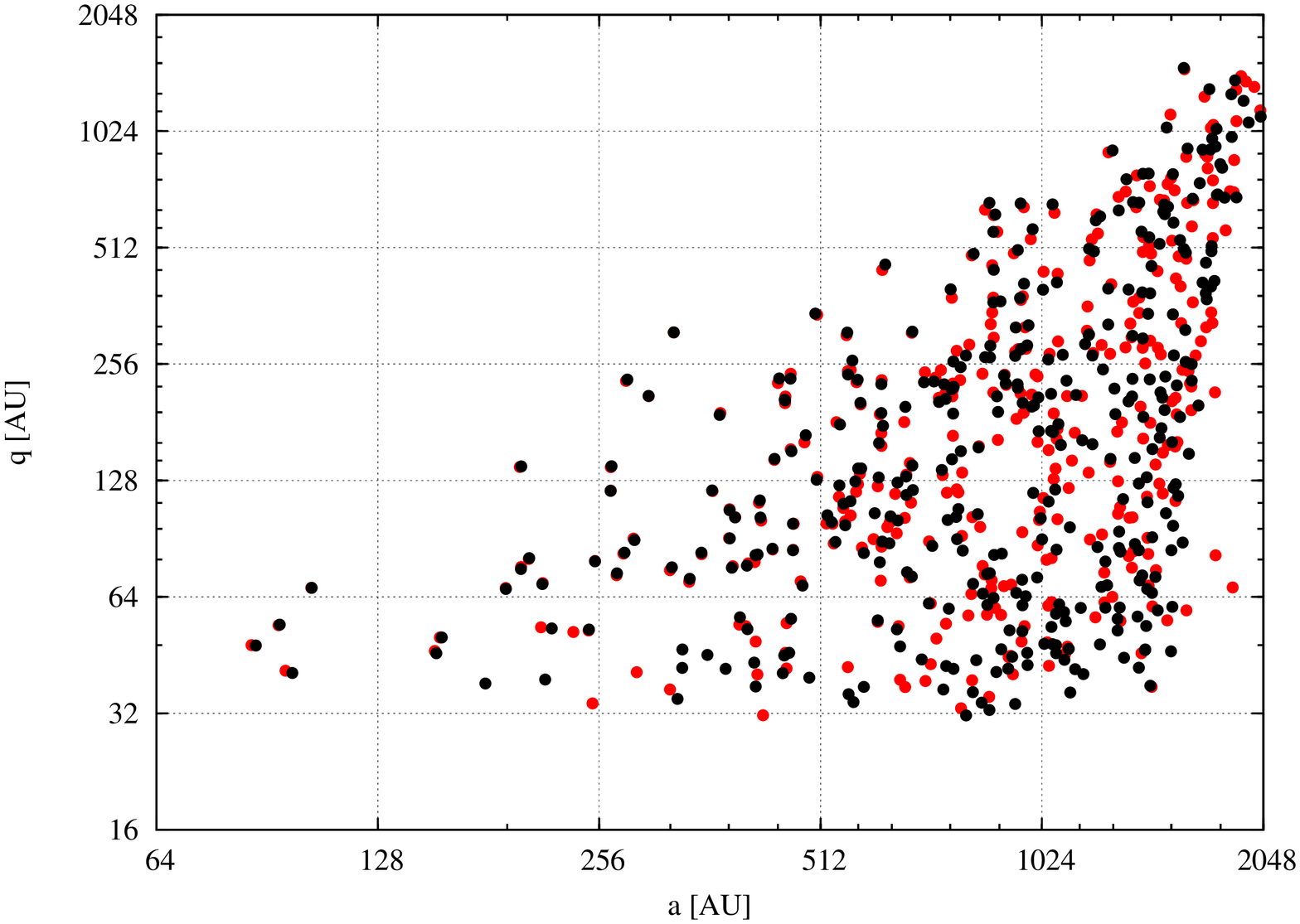}}
\caption{The inner Oort cloud in semi-major axis-perihelion space. The black dots are the positions of all IOC objects before
migration. The red (in the online version) dots are after 4-planet giant migration and subsequent evolution to the current age of the 
Solar System. It is clear that little orbital evolution has occurred.}
\label{ioc}
\end{figure}

Here we tried to determine what happens to the IOC under the influence of four planet giant planet migration and subsequent evolution
of the Solar System. We subjected the IOC to the same evolution of giant planet migration, we ran sixteen simulations (the 
free parameters are Hernquist or Plummer potential, gas removal time, central concentration) and we focused on the inner core, which is
defined as objects with semi-major axis $a<1000$~AU; Sedna and 2012 VP113 are within this region. The outcome of one simulation is
depicted in Fig.~\ref{ioc}, which came from a Hernquist potential run, and it is representative for the outcome of all the other 
simulations. We depict the IOC in semi-major axis-perihelion space, as was done in Brasser et al. (2006, 2012). The black bullets 
denote the IOC before giant planet migration; this is reminiscent of the state just after the Sun had left the cluster. In contrast 
the red bullets denote the IOC after migration and subsequent evolution to the current age of the Solar System.
It is evident from the figure that the structure of the IOC remained virtually unchanged. Indeed, for most objects the orbital energy
and perihelion distance has barley evolved. We repeat that our starting IOC contained no objects with $q<35$~AU. We anticipated that
Neptune's migration and the subsequent evolution would clear away any objects with original perihelion $q<40$~AU, but when we
computed the final perihelion distribution this turned out not to be the case. Some objects underwent distant encounters with Neptune
and were subsequently scattered onto Centaur orbits or ejected from the Solar System, but most objects with $q\gtrsim 35$~AU underwent
little evolution. From this we conclude that, in this scenario for the migration of the giant planets, the inner core of the IOC i.e.
the region inside 2\,000~AU, remained intact and objects that were placed there in the early stages of the Solar System are still there
today. 

\begin{figure}
\resizebox{\hsize}{!}{\includegraphics[angle=-90]{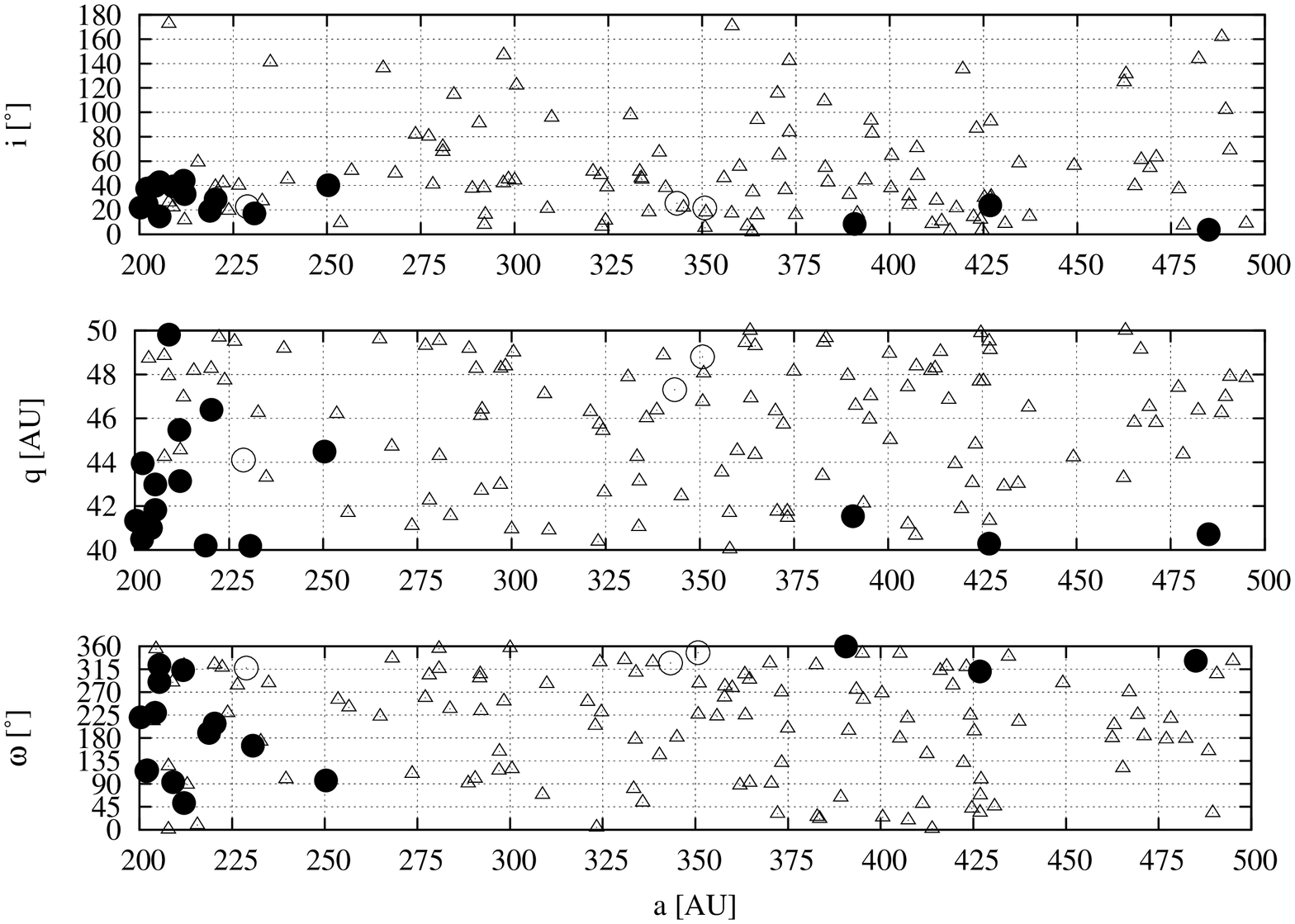}}
\caption{Inclination (top), perihelion distance (middle), and argument of perihelion of detached objects ($q>40$~AU) as a 
function of their semi-major axis. Open circles represent the orbits known detached KBOs excluding Sedna and 2012 VP113. Filled circles
mark the orbits of detached KBOs produced in the ESD simulations. Open triangles represent the orbits produced in the simulations of
the IOC.}
\label{cr105}
\end{figure}
We summarise the findings of this study in Fig.~\ref{cr105} below. This is a three-panel plot highlighting the region where the
origin of objects is somewhat ambiguous. We selected the region with semi-major axes between 200~AU and 500~AU and perihelion distances
between 40~AU and 50~AU. The top panel depicts the inclination versus semi-major axis, the middle panel perihelion versus semi-major
axis, and the bottom panel plots argument of perihelion versus semi-major axis. The triangles are IOC objects, the bullets are
simulated ESD objects from the simulations of Brasser \& Morbidelli (2013), while the open circles are real observed objects. We have
plotted many more IOC objects as there are ESD objects. This is intentional and reflects the relative probabilities of objects
ending up in this region through an ESD and IOC origin. For semi-major axes shorter than 250~AU there are plenty of ESD objects,
including a few outliers beyond 350~AU. However, when $a>250$~AU no ESD objects have $q>45$~AU. The inclination of most ESD objects is
low, between 20$^\circ$ and 30$^\circ$. It appears that the IOC objects have a prograde excess and most IOC objects have inclinations
lower than 60$^\circ$. This is expected because the inclination distribution is not made isotropic by the Galactic tide and passing
stars, and the inner edge of the IOC is known to be flattened (Brasser et al., 2006, 2012; Kaib \& Quinn, 2008).

\subsection{Formation efficiency and argument of perihelion distribution}
From the above results it appears that objects in the third region ($a>250$~AU, $q>40$~AU) should have an IOC origin rather than an SD
one. We can verify this claim by calculating the probability of placing objects in this region based on our numerical simulations. We
first discuss the SD.

From the simulations of Brasser \& Morbidelli (2013) we determined that in steady state only 7.5\% of SD objects have $a>250$~AU. Of
these 0.15\% have both $a>250$~AU and $q>39$~AU. Our additional simulations show that approximately 0.2\% of all simulated objects
reached orbits with $42<q<45$~AU and $250<a<1000$~AU. Thus, in summary, the overall probability of producing objects with $42<q<45$~AU
and $250<a<1000$~AU is $3 \times 10^{-6}$. Combining the two and scaling by the total number of integrated particles, we calculate 
that the probability of producing objects with $250<a<1000$~AU and $q>45$~AU by this mechanism is lower than $4 \times 10^{-8}$, and is 
the reason we saw no such objects in our simulations.

For an IOC origin we proceed as follows. The probability of placing an IOC object with $250<a<1000$~AU and $q>45$~AU is approximately
0.1\%, using a total IOC formation probability of 1.5\% (Brasser et al., 2012). When we restrict ourselves to imposing that the
perihelion distance $42<q<45$~AU the probability is lowered to $2 \times 10^{-5}$. This is at least an order of magnitude higher than
an ESD origin, and thus it is likely that almost all objects with $a>250$~AU and $q>42$~AU are IOC objects.

We now turn our attention to the bottom panel of Fig.~\ref{cr105} where we plotted the instantaneous argument of perihelion of
real and simulated ESD and IOC objects versus their semi-major axis. Trujillo \& Sheppard (2014) suggested that ESD and IOC
objects displayed a grouping in $\omega$ around 0$^\circ$, and they attributed this to the influence of an unseen planet residing in
the IOC. Trujillo \& Sheppard (2014) included objects with semi-major axis $a<250$~AU and perihelion $q<40$~AU, which, as we have
shown, are clearly dominated by interaction with Neptune. Indeed, we advocate that there are to date only four IOC objects discovered:
Sedna, 2004 VN112, 2010 GB174 and 2012 VP113. While it is true that all four of these objects have $\omega$ fairly close to 0$^\circ$,
we do not think that these objects are sufficient in number to claim that the $\omega$ clustering is real or just a coincidence. In
fact, with our current model of migration, we rule out such clustering being real for objects with $a>250$~AU because they just undergo
precession at a rate the scales as $a^{-7/2}$. The precession period of Sedna is of the order of 1~Gyr (Brasser et al., 2006) and thus
any initial grouping that was present when the Sun left the star cluster is long gone. In summary, more objects are needed to confirm
or deny whether this grouping in $\omega$ near 0$^\circ$ is real.

\section{Five-planet model}
In the above sections we examined the influence of giant planet migration, using a four planet model (see \ref{sec:4planet}), on the 
ESD and the IOC. We determined that the SD ends abruptly at $q=40$~AU when $a>250$~AU and that objects with $a>250$~AU and $q>40$~AU
most likely have an IOC origin. Does the five planet model change this picture?

An indication of the possible success or failure of the five planet model comes from Gladman \& Chan (2006), who reported that an
Earth-mass rogue planet scattering off Neptune with a dynamical lifetime of 160~Myr was able to lift the perihelion of many SD objects
into the ESD and IOC region. The time scale for doing so is approximately
\begin{equation}
 P_{\rm sec} \sim 3640\Bigl(\frac{m_p}{m_\oplus}\Bigr)^{-1}\Bigl(\frac{q_p}{5\,{\rm AU}}\Bigr)^{3/2}(1+e_p)^{3/2}\quad \rm{kyr},
\end{equation}
where $P_{\rm sec}$ is the secular time scale, $m_p$, $q_p$, and $e_p$ are the mass, perihelion distance and
eccentricity of the rogue planet. For an Earth mass planet with perihelion $q_p \sim 30$~AU and $e_p \sim 0.85$ we have $P_{\rm sec}
\sim 110$~Myr, comparable to the dynamical lifetime of the rogue. However, for an ice giant of Neptune's mass scattering off Jupiter,
we have $e_p \sim 1$ and $P_{\rm sec} \sim 0.6$~Myr. The typical dynamical lifetime of the ice giant on an eccentric orbit with
perihelion near Jupiter is about 50~kyr, an order of magnitude shorter, suggesting that the eccentric ice giant may not account for
much influence on the SD and the IOC. 

\begin{figure}
\resizebox{\hsize}{!}{\includegraphics[angle=-90]{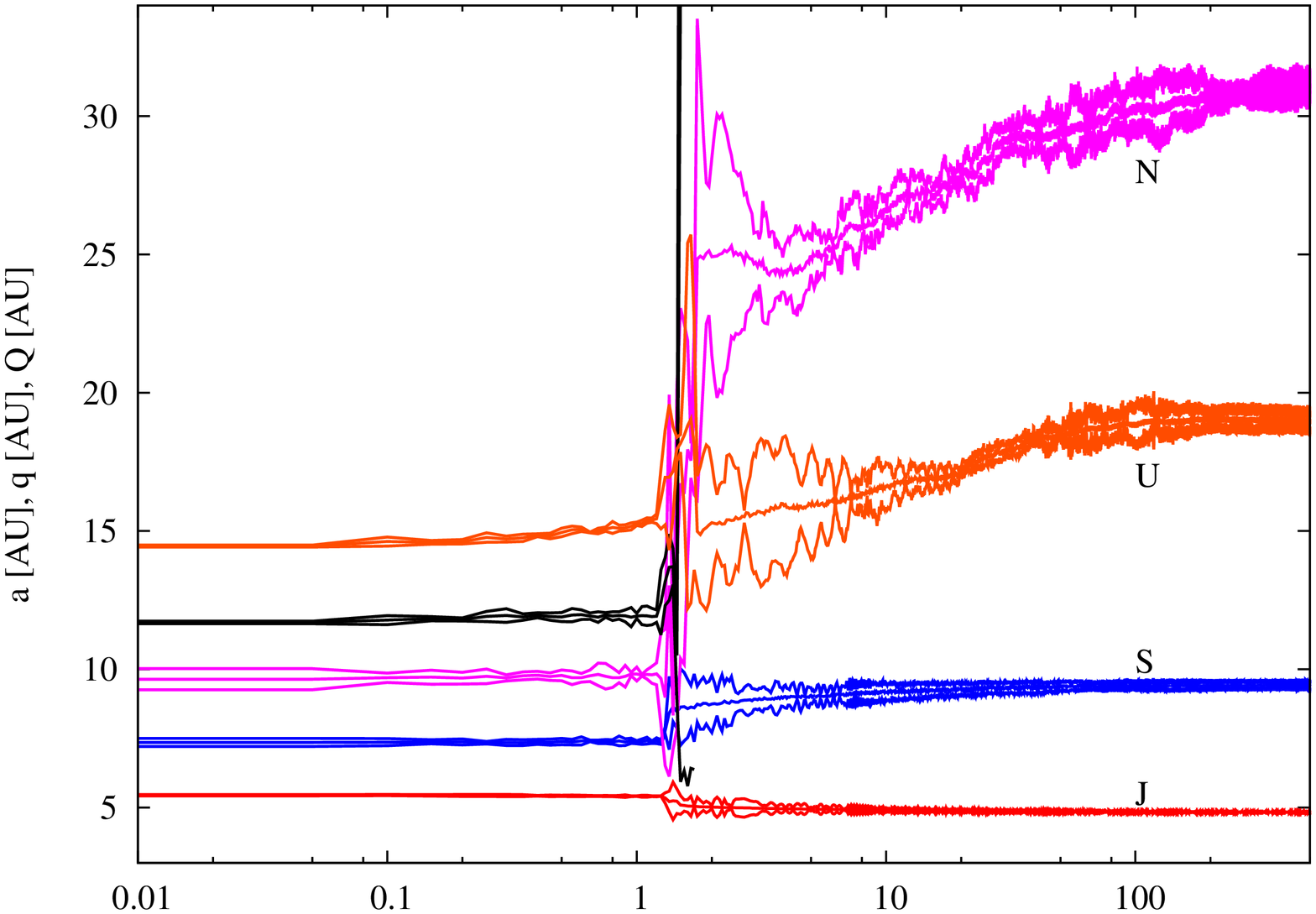}}
\caption{Evolution of the perihelion, semi-major axis and aphelion of a 5-planet migration simulation. The second ice giant is ejected 
by Jupiter after 2 Myr from the beginning of the simulation.}
\label{5plevo}
\end{figure}

To test the influence of the fifth planet on the SD and the IOC we ran additional simulations. The initial conditions for the giant 
planets and the SD were taken from Brasser \& Lee (2014). The five planets were placed in the 3:2, 3:2, 4:3, 4:3 resonant chain. The 
planetesimal disk consisted of 2\,000 planetesimals with a surface density that scales with heliocentric distance as $\Sigma \propto 
r^{-1}$. The planetesimals were included to damp the random velocities of the giant planets and to induce migration. The outer edge
of the disk was always set at 30~AU. The inner edge was $\Delta = 0.5$~AU from the outermost ice giant and the total disc mass was
35$M_{\oplus}$. We generated a hundred different realisations of the same initial conditions by giving a random deviation of
$10^{-6}$~AU to the semi-major axis of each planetesimal. The simulations were performed with the SyMBA integrator (Duncan et al.,
1998), with a time step of 0.35~yr. Particles were removed farther than 2000~AU from the Sun, when they hit a planet or came closer
than 1~AU to the Sun, and the planets' evolution was output every 100~yr. The simulations ran for 500~Myr. An example of the evolution 
that produces the current Solar System rather well is depicted in Fig.~\ref{5plevo}.

\begin{figure}
\resizebox{\hsize}{!}{\includegraphics[angle=-90]{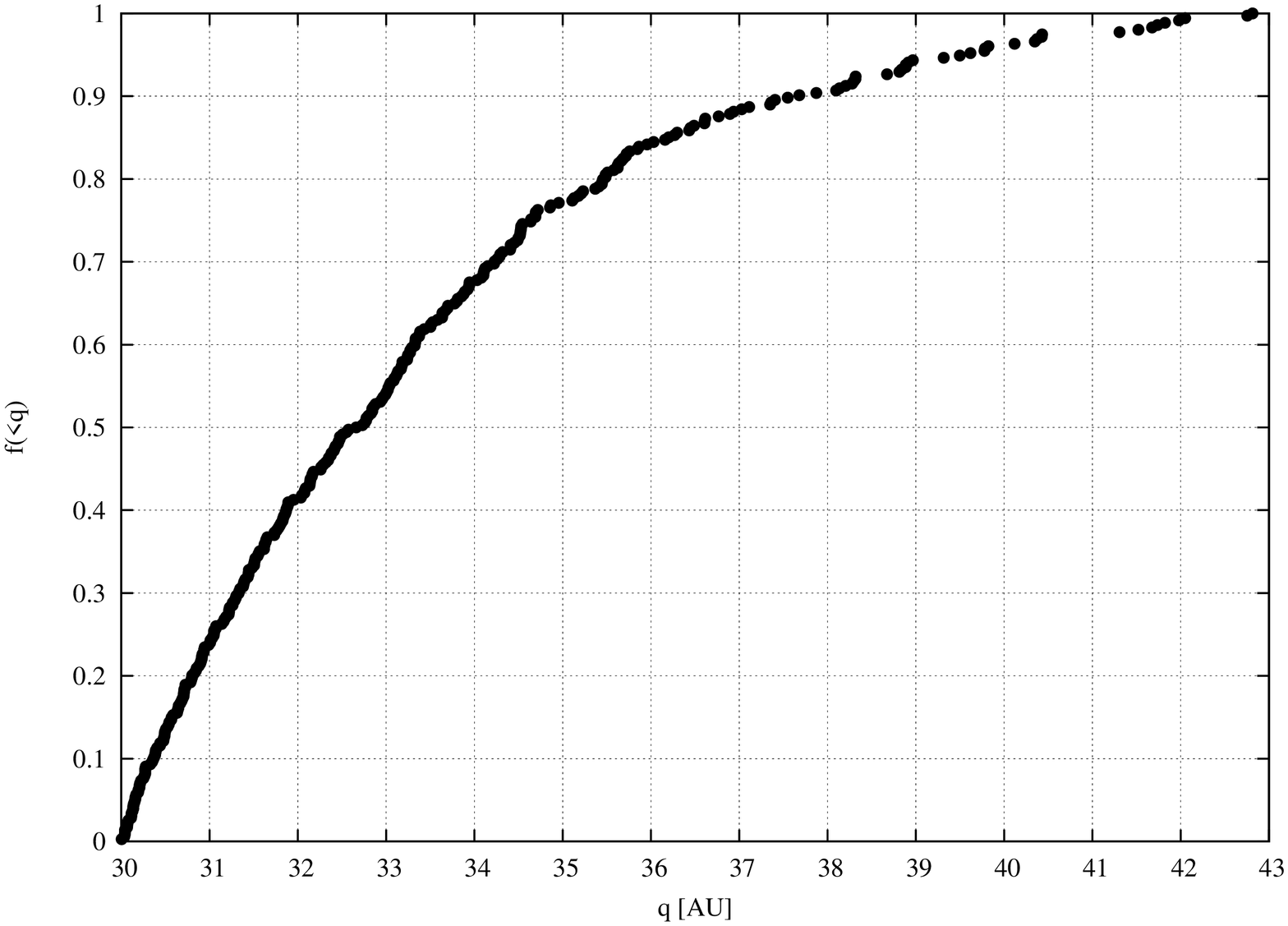}}
\caption{Cumulative distribution (solid line) of the perihelion distance of Scattered Disc objects with semi-major axis $a>250$~AU for 
the 5-planet model after 500 Myr of evolution.}
\label{5plcumq}
\end{figure}

We analysed the evolution of the planets to determine how long the 5th planet crossed the Kuiper Belt and SD.  We 
report that this is consistently shorter than 30~kyr. With this in mind we performed a further 100 simulations but this time we used 
12\,000 with the same initial conditions as the runs described above. We then analysed the distribution of planetesimals in those 
simulations where the planets were within 10\% of their current orbits. We did not care about the final eccentricities or secular 
architecture. In the end 23 runs matched this constraint, so that we effectively integrate 276\,000 planetesimals. We analysed the 
cumulative perihelion distribution of all planetesimals with $a>250$~AU and combined them into Fig.~\ref{5plcumq}. Note that as in the 
4-planet case there are no objects with $q\gtrsim 42$~AU.

Gladman \& Chan (2006) were able to produce ESD objects through the secular interaction of SD objects with their rogue planet. Our
results differ from theirs because the cumulative time the fifth planet spends on an eccentric orbit is much shorter than the secular
time scale calculated above. We report that the IOC remains virtually unchanged - just like in the four planet case. The 
fifth planet did not place any of the 276\,000 planetesimals from the SD into the ESD, thus the probability of the fifth planet lifting 
a planetesimal into the ESD is lower than $3 \times 10^{-6}$. In comparison, in the previous section we determined that an IOC origin 
yields a probability that is an order of magnitude higher ($2 \times 10^{-5}$). After the fifth planet was ejected we recorded no 
objects with $q>43$~AU. Thus we conclude that the influence of an ejected  fifth giant planet on the outer Solar System during the 
ejection stage is negligible with the final orbital distribution of the ED and IOC virtually indistinguishable from the four planet 
model. In addition, this model cannot explain the possible clustering in the argument of perihelion either.

\section{Conclusions}
We have examined the Inner Oort cloud produced by a stellar cluster birth origin after including the impact of Neptune's migration and
the effects of the Solar System's subsequent evolution over 4 billion years. We explored models with four and five giant planets.
We also explored the extent of the ESD, examining the origin of objects with semi-major axes and perihelion less than Sedna and 2012
VP113. We find that for a four-planet migration model an object can be classified as belonging to the IOC if it has a semi-major axis
larger than 250~AU and a perihelion beyond 45~AU. For objects with perihelia detached from Neptune, but lower semi-major axes, their
origin is ambiguous and cannot be used to constrain the formation of the IOC. The five planet models do not change the outcome. Further
wide-field surveys and discoveries of new potential IOC objects with semi-major axes beyond 250~AU are needed to further test this
origin scenario.

Interestingly Trujillo \& Sheppard (2014) find an apparent correlation of argument of perihelion for the two confirmed members of the
IOC and detected KBOs objects closer in to Neptune. The question is what mechanism could be responsible for this argument of perihelion
correlation because the orbits of objects in the IOC and ESD undergo precession from Neptune that does not automatically produce this
orbital correlation. From simple dynamical modelling Trujillo \& Sheppard (2014) suggest the apparent correlation could be the result
of perturbations from an unseen planet. However, since in the framework of the planet migration model we employed it appears there are
to date only four confirmed IOC members. More discoveries with $q>45$~AU and $a>250$~AU are needed to confirm or reject this
clustering in the argument of perihelion.

\section*{Acknowledgements}
MES is supported in part by an Academia Sinica Postdoctoral Fellowship. We thank Rodney Gomes, Hal Levison and Alessandro Morbidelli 
for simulating discussions, and Nathan Kaib for his review. This work was made possible through the use of the ASIAA HTCondor computer 
clusters at ASIAA. The HTCondor software program was developed by the HTCondor Team at the Computer Sciences Department of the 
University of Wisconsin-Madison. All rights, title, and interest in HTCondor are owned by the HTCondor Team.

\section*{Bibliography}
Batygin K., Brown M.~E., Fraser W.~C., 2011, ApJ, 738, 13 \\
Batygin K., Brown M.~E., Betts H., 2012, ApJ, 744, L3 \\
Becker A.~C., et al., 2008, ApJ, 682, L53 \\
Brasser R., Duncan M.~J., Levison H.~F., 2006, Icar, 184, 59\\
Brasser R., Duncan M.~J., Levison H.~F., Schwamb M.~E., Brown M.~E., 2012, Icar, 217, 1\\
Brasser R., Morbidelli A., 2013, Icar, 225, 40\\
Brasser, R., Lee, M. H., 2014, AJ, submitted \\
Brown, M.~E., Trujillo, C., Rabinowitz, D.\ 2004.\ ApJ 617, 645 \\
Chen Y.-T., et al., 2013, ApJ, 775, L8 \\
Duncan, M.~J., Levison, H.~F.\ 1997.\ Science 276, 1670 \\
Gladman B., Holman M., Grav T., Kavelaars J., Nicholson P., Aksnes K., Petit J.-M., 2002, Icar, 157, 269\\
Gladman B., Chan C., 2006, ApJ, 643, L135\\
Gladman, B., Marsden, B.~G., Vanlaerhoven, C.\ 2008.\ The Solar System Beyond Neptune 43-57, University of Arizona Press. Tucson, AZ,
USA.\\ 
Gomes R.~S., Gallardo T., Fern{\'a}ndez J.~A., Brunini A., 2005, CeMDA, 91, 109\\
Gomes R.~S., Matese J.~J., Lissauer J.~J., 2006, Icar, 184, 589\\
Gomes R.~S., 2011, Icar, 215, 661\\
Heisler, J., Tremaine, S., Alcock, C.\ 1987.\ Icarus 70, 269 \\
Hernquist L., 1990, ApJ, 356, 359 \\
Hills J.~G., 1981, AJ, 86, 1730\\
Holmberg, J., Flynn, C.\ 2000.\ MNRAS 313, 209 \\
Kaib, N.~A., Quinn, T.\ 2008. Icarus 197, 221 \\
Kaib, N.~A., Quinn, T., Brasser, R.\ 2011.\ AJ 141, 3 \\
Kenyon S.~J., Bromley B.~C., 2004, Natur, 432, 598\\
Lada C.~J., Lada E.~A., 2003, ARA\&A, 41, 57\\
Levison, H.~F., Duncan, M.~J.\ 1994.\ Icarus 108, 18 \\
Levison, H.~F., Dones, L., Duncan, M.~J.\ 2001.\ AJ 121, 2253. \\
Levison H.~F., Morbidelli A., Van Laerhoven C., Gomes R., Tsiganis K., 2008, Icar, 196, 258\\
Lykawka P.~S., Mukai T., 2007, Icar, 189, 213\\
McMillan, P.~J., Binney, J.~J.\ 2010.\ MNRAS 402, 934 \\
Morbidelli A., Levison H.~F., 2004, AJ, 128, 2564 
Morbidelli A., Brasser R., Gomes R., Levison H.~F., Tsiganis K., 2010, AJ, 140, 1391\\
Nesvorn{\'y} D., 2011, ApJ, 742, L22\\
Nesvorn{\'y} D., Morbidelli A., 2012, AJ, 144, 117\\
Ram{\'{\i}}rez I., et al., 2014, ApJ, 787, 154\\
Schwamb, M.~E., Brown, M.~E., Rabinowitz, D.~L., Ragozzine, D.\ 2010.\ ApJ 720, 1691 \\
Trujillo C.~A., Sheppard S.~S., 2014, Natur, 507, 471
\end{document}